\documentclass[letterpaper]{article}
\usepackage[preprint]{aaai2027}
\usepackage[hyphens]{url}
\usepackage{graphicx}
\usepackage{natbib}
\usepackage{caption}
\usepackage{booktabs}
\usepackage{array}
\usepackage{amsmath}
\usepackage{amssymb}
\title{Auto Research for Materials:\\Auditable AI-Scientist Workflows with Held-Out Transfer}
\author{
Jingjie Ning\textsuperscript{\rm 1},
Xiaochuan Li\textsuperscript{\rm 1},
Shanshan Zhong\textsuperscript{\rm 1},
Ji Zeng\textsuperscript{\rm 1},
Guolin Ke\textsuperscript{\rm 2}
}
\affiliations{
\textsuperscript{\rm 1}School of Computer Science, Carnegie Mellon University\\
\textsuperscript{\rm 2}DP Technology\\
\{jening, xiaochu4, szhong2, jizeng\}@cs.cmu.edu, kegl@dp.tech
}

\begin{document}
\maketitle

\begin{abstract}
Auto Research uses language-model agents to propose, implement, and evaluate machine-learning changes in a closed loop, but is usually judged by its terminal pipeline. A terminal score cannot reveal which technical decision produced a gain or distinguish a reusable discovery from a change adapted to development feedback. We introduce intervention-centered Auto Research, which validates research decisions rather than only final artifacts and makes their reliability measurable. Feature, Model, Representation, and Data axes are searched independently with inner five-fold feedback. Each axis winner is frozen before an outer-holdout matrix compares all alternatives on evidence the loop never sees. Across 701 agent-executed attempts spanning ten Matbench endpoints, outer evidence confirms the selected intervention on nine of ten endpoints and preserves 89.3\% of non-tied intervention orderings. It also rejects an aggregate Representation gain that inner feedback endorsed. The resulting matrix reveals an information-dependent hierarchy. Composition-only tasks support several routes to improvement, whereas structure-informed tasks favor local geometry features and complementary tree ensembles. A subsequent compatibility test combines already frozen Feature and Model code without further search or tuning and raises mean outer-holdout improvement from 19.0\% to 26.3\%. By validating decisions rather than only artifacts, this design turns adaptive search into reusable evidence wherever agents propose executable alternatives against a fixed evaluator.
\end{abstract}

\section{Introduction}

Auto Research uses language-model agents to propose, implement, evaluate, and revise machine-learning changes in a closed loop. Existing benchmarks and AI-scientist studies establish that these agents can improve programs, task scores, and research artifacts \citep{huang2024mlagentbench,chan2025mlebench,jiang2025aide,lu2026ai_scientist,novikov2025alphaevolve,martinek2026agentomics,gao2026autoscientists}. The next question is whether decisions from adaptive search remain supported after its feedback loop ends. A final pipeline may entangle data, representation, feature, and model changes, so its score cannot identify what carried the gain. Repeated selection against the same feedback can also favor a change that does not survive new data \citep{cawley2010selection,dwork2015reusable,recht2019imagenet}. Decision-level validation matters for cumulative science because later work builds on the change that remains justified and reusable, and a high-scoring artifact alone does not identify that change. Auto Research therefore needs a methodology that preserves attribution during discovery and separates search feedback from final validation.

\begin{figure*}[t]
\centering
\includegraphics[width=\textwidth]{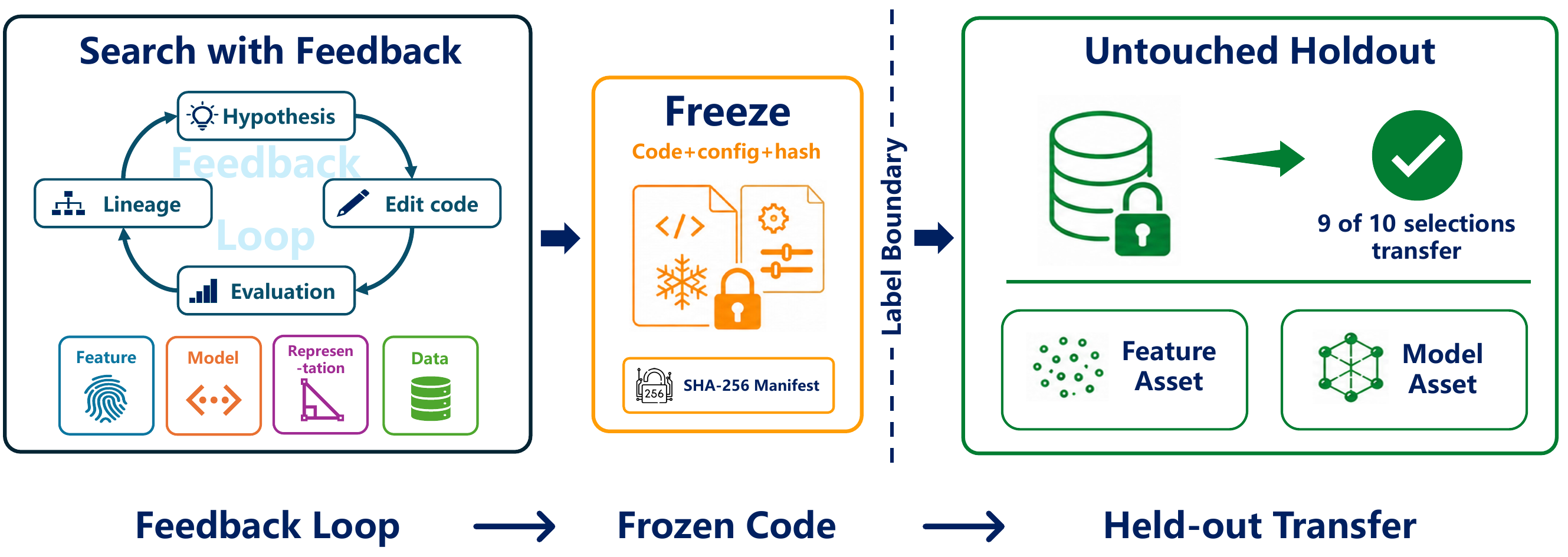}
\caption{Overview of intervention-centered Auto Research. Independent Feature, Model, Representation, and Data searches use inner five-fold feedback. At the label boundary, the selected code, configuration, and SHA-256 manifest are frozen before the untouched outer holdout is evaluated; 9 of 10 inner-selected choices remain best outside the loop.}
\label{fig:overview}
\end{figure*}

We develop intervention-centered Auto Research. A \emph{research intervention} is an executable modification confined to one declared axis of a research workflow. We use \emph{axis} for the editable category, \emph{attempt} for one evaluated proposal, and \emph{intervention} for the selected frozen code. In this study, the axes are Feature, Model, Representation, and Data. The agent proposes and implements candidates on each axis under equal inner feedback. An outer-holdout matrix then compares every intervention on selection regret, ordering, breadth, and compatibility. This makes research decisions attributable and directly comparable where a terminal-pipeline score cannot.

The materials results show both sides of decision validation. Inner feedback retains a structure Representation intervention with a 1.4\% gain, but the outer holdout rejects it with a 3.0\% loss. Independently searched structure Feature and Model interventions instead lead on different property families. A later deterministic compatibility test combines their frozen code without further search or tuning and improves all six structure endpoints. A terminal score alone cannot expose either the sign reversal or the reusable division of labor.

Materials prediction makes information availability an explicit design factor. Composition-only endpoints expose formula evidence, whereas structure-informed endpoints also expose atomic coordinates. We compare axes within each regime under matched search and evaluation contracts, then use a controlled input comparison on fixed endpoints to separate information type from input dimension. This tests whether an autonomous scientist directs effort toward missing information or simply repeats a preferred modeling move.

Across seven campaigns and ten Matbench endpoints, inner five-fold feedback selects the best intervention on nine outer holdouts, while 89.3\% of non-tied pairs retain their ordering. Composition supports several axes. With structure inputs, Feature leads on phonons and perovskites, while Model leads on elastic and optical tasks. Combining independently found Feature and Model interventions raises mean outer-holdout improvement from 19.0\% to 26.3\%.

This paper makes three contributions.
\begin{itemize}
    \item \textbf{Intervention-centered Auto Research.} We combine axis-isolated discovery, inner five-fold feedback, code freezing, and once-only outer-holdout evaluation. The resulting design converts open-ended search into attributable interventions with measurable selection, ordering, breadth, and compatibility.
    \item \textbf{Evidence for held-out transfer.} Across 701 attempts and ten endpoints, inner five-fold feedback selects the best tested choice on nine outer holdouts, with 0.228 percentage-point mean regret and 89.3\% pairwise agreement. The same audit rejects an apparent Representation gain that inner feedback alone would have endorsed.
    \item \textbf{Information-dependent findings.} Composition-only tasks support several useful axes. With crystal structure, geometry features and complementary tree models lead among those tested, while added composition embeddings do not replace structural information. Separately found feature and model code remains compatible and improves all six structure outer holdouts.
\end{itemize}

\section{Related Work}

\paragraph{Automated pipeline and program search.}
Auto-sklearn and TPOT search machine-learning pipelines, AutoGluon builds stacked ensembles, and AutoML-Zero evolves learning programs from basic operations \citep{feurer2015autosklearn,olson2016tpot,erickson2020autogluon,real2020automlzero}. FunSearch, AIDE, and AlphaEvolve use language models to search executable programs \citep{romeraparedes2024funsearch,jiang2025aide,novikov2025alphaevolve}. Controlled decompositions of multi-model pipelines further show that an aggregate gain can combine distinct mechanisms \citep{ning2026revision}. We ask how adaptive research can preserve attribution and test each technical change after search.

\paragraph{Evaluation of AI scientists.}
The AI Scientist automates a research cycle from ideation through manuscript review \citep{lu2026ai_scientist}. MLAgentBench and MLE-bench evaluate terminal solutions, while SkillLearnBench separates skill, trajectory, and outcome quality and finds that feedback gains depend on task structure \citep{huang2024mlagentbench,chan2025mlebench,zhong2026skilllearnbench}. Agentomics and AutoScientists emphasize validated artifacts or long-running team search \citep{martinek2026agentomics,gao2026autoscientists}. These evaluations ask what an agent can produce. We ask whether frozen interventions remain supported outside the loop.

\paragraph{The closest Auto Research studies.}
An anonymized prior study used lineage to search fixed-evaluator training recipes \citep{anonymous2026recipes}. A second study certified one axis-isolated molecular-prediction intervention against its baseline \citep{anonymous2026molecules}. The former evaluates a terminal recipe, and the latter evaluates one selected change against its baseline. We instead freeze every independently searched axis winner and compare them under shared outer evidence. This produces a different evidence object that measures decision regret, alternative ordering, reusable breadth, and cross-axis compatibility rather than another winner-versus-baseline certificate.

\paragraph{Autonomous materials research.}
LLMatDesign and SparksMatter use language-model feedback to propose and refine candidate materials, while MASTER and MADE optimize sequential simulation or candidate-selection campaigns \citep{jia2024llmatdesign,ghafarollahi2026sparksmatter,rothfarb2026master,malik2026made}. AutoMat instead tests whether coding agents can reproduce findings from materials papers \citep{huang2026automat}. Autonomous laboratories couple AI with physical experiments \citep{szymanski2023alab,boiko2023coscientist}. Their principal outputs are candidates, discovery efficiency, experimental execution, or reproduction success. We hold the materials tasks and evaluator fixed and study controlled changes to a predictive pipeline. This design exposes which intervention type remains useful under composition-only and structure-informed inputs.

Table~\ref{tab:positioning} positions this work by the evidence each study family produces after search. A strong final artifact does not reveal whether its constituent changes survive separately. Our design preserves the full set of frozen alternatives and evaluates which technical results transfer beyond adaptive feedback.

\begin{table*}[t]
\centering
\small
\setlength{\tabcolsep}{8pt}
\renewcommand{\arraystretch}{1.10}
\begin{tabular}{@{}>{\raggedright\arraybackslash}p{0.18\textwidth}
                    >{\raggedright\arraybackslash}p{0.31\textwidth}
                    >{\raggedright\arraybackslash}p{0.41\textwidth}@{}}
\toprule
\textbf{Study family} & \textbf{Primary question} & \textbf{Unit evaluated after search} \\
\midrule
Program and pipeline search & Did search find a strong program? & Terminal program or pipeline; final task score \\
End-to-end AI scientists & Did the agent produce a valid artifact? & Paper or artifact; execution or review \\
Materials agents & Did the loop find a useful candidate? & Candidate or campaign; simulation or synthesis \\
Prior axis certification & Did one selected change generalize? & Frozen intervention against its baseline \\
\textbf{This work} & \textbf{Which independently searched intervention remains useful under each information regime?} & \textbf{All frozen alternatives; outer-holdout selection, ordering, breadth, and compatibility} \\
\bottomrule
\end{tabular}
\caption{Positioning by the evidence produced after search. This work compares all independently searched axes on selection, ordering, breadth, and compatibility.}
\label{tab:positioning}
\end{table*}

\paragraph{Materials prediction and evaluation.}
Matbench standardizes tasks and folds for inorganic property prediction \citep{dunn2020matbench}; Matbench Discovery extends evaluation to crystal-stability prediction under realistic discovery objectives \citep{riebesell2025matbench_discovery}. Composition descriptors and models include Magpie, MODNet, Roost, and CrabNet \citep{ward2016magpie,debreuck2021modnet,goodall2020roost,wang2021crabnet}, while CGCNN, MEGNet, coNGN, and ALIGNN use crystal geometry \citep{xie2018cgcnn,chen2019megnet,congn,choudhary2021alignn}. We use these families to define candidate interventions rather than proposing another fixed architecture.

\section{Intervention-Centered Auto Research}

\subsection{Evidence-Separated Search}

The method separates discovery, selection, and validation. A \emph{campaign} is a sequence of propose, edit, and evaluate attempts on one editable axis using only inner five-fold feedback. It freezes the exact selected code as an intervention. The same feedback chooses among these interventions and the baseline. A once-only \emph{outer-holdout matrix} then scores every alternative on evidence the loop never saw.

Let $\mathcal{T}$ be the prediction endpoints and let $\mathcal{A}$ contain the editable axes. Our primary outcome is the outer-holdout regret of the intervention selected by inner five-fold feedback, not the gain of the terminal program. For inner fold $k$, candidate $c$ has direction-corrected normalized improvement
\begin{equation}
u_{t,k}(c)=
\begin{cases}
(b_{t,k}-s_{t,k}(c))/b_{t,k}, & \text{for MAE},\\
(s_{t,k}(c)-b_{t,k})/b_{t,k}, & \text{for ROC-AUC},
\end{cases}
\end{equation}
where $b_{t,k}$ and $s_{t,k}(c)$ are the baseline and candidate scores on the same fold. During search, every score is computed only inside the outer training set. We use $K=5$ fixed folds and return
\begin{equation}
d_t(c)=\frac{1}{K}\sum_{k=1}^{K}u_{t,k}(c)
\end{equation}
to the agent. Thus MAE and ROC-AUC enter cross-endpoint aggregation on the same ``larger is better'' scale. Aggregation therefore reflects relative improvement rather than the raw magnitude of each endpoint's metric. The per-endpoint choice still uses only that endpoint's $d_t$. The five-fold average reduces dependence on any single partition and gives every candidate the same multi-partition comparison.

\paragraph{Why inner five-fold feedback matters.}
In a closed-loop search, feedback both scores the current candidate and determines which hypothesis the agent extends next, so a split-specific fluctuation can redirect many later experiments. Every candidate receives the same five inner partitions and the same evidence budget before the record is updated. No single partition controls the next branch. The outer holdout has a different role and is evaluated only after code and choices have been frozen. Combining multi-partition feedback with a code freeze and once-only outer evaluation gives continuous Auto Research a stable ranking signal and a separate audit of the resulting decision.

Each axis $a$ has a separate campaign and set of editable files $\mathcal{C}_a$. The campaign freezes the intervention
\begin{equation}
c_a^*=\arg\max_{c\in\mathcal{C}_a}
\frac{1}{|\mathcal{T}_a|}\sum_{t\in\mathcal{T}_a}d_t(c).
\end{equation}
This produces one frozen intervention per searched axis. For each endpoint, inner five-fold feedback then chooses among the baseline and these interventions,
\begin{equation}
\hat a_t=\arg\max_{a\in\mathcal{A}\cup\{0\}}d_t(c_a^*),
\end{equation}
with $c_0$ denoting the baseline. The outer-holdout labels play no role in either selection.

\subsection{Materials Instantiation}

The \textbf{Feature} axis changes hand-crafted descriptors derived from composition or structure. The \textbf{Model} axis changes the estimator, ensemble, regularization, or output calibration. The \textbf{Representation} axis changes embedding-based inputs, including fraction-weighted mat2vec, MEGNet-derived, and one-hot elemental representations \citep{tshitoyan2019mat2vec,chen2019megnet}. The \textbf{Data} axis changes agent-curated candidate training rows after formula and source checks. Every campaign starts from the same baseline. Attempts may build on earlier retained code within that campaign, while hashes verify that files outside its assigned axis remain unchanged. Axis isolation is important because an end-to-end winner that changes all four axes cannot show which research decision carried the gain.

Each axis also represents a different diagnosis of prediction error. Feature search asks whether the current variables omit useful physics. Representation search asks whether known inputs are encoded poorly. Model search changes the inductive bias that maps inputs to targets. Data search asks whether the observed examples are sufficient. Keeping these diagnoses separate means that a winning axis identifies where the next unit of research effort should go instead of only recording a better configuration.

Within a campaign, the agent proposes a hypothesis, edits the permitted files, invokes the evaluator, and receives the five-fold mean plus execution status. Earlier hypotheses, diffs, scores, and failed experiments are available when proposing the next attempt. This experiment record guides search but is not an evaluation target.

\subsection{Frozen-Code Held-Out Transfer}

After freezing all $c_a^*$, the trusted evaluator fits each intervention on outer-training data and scores the holdout. Agent-written code receives training labels and unlabeled evaluation inputs; only the evaluator process can read outer-holdout labels.

Let $h_t(c)$ be the direction-corrected normalized outer-holdout improvement, defined in the same way as Eq.~(1). We measure the cost of choosing from inner five-fold feedback with \emph{selection regret}
\begin{equation}
r_t=\max_{a\in\mathcal{A}\cup\{0\}}h_t(c_a^*)-h_t(c_{\hat a_t}^*).
\end{equation}
An endpoint exhibits \emph{held-out transfer} when $r_t=0$. This comparison includes the baseline and independently searched single-axis interventions, but not later combinations. We also compare every within-task pair of interventions. Pairwise ordering agreement asks whether the sign of $d_t(c_a^*)-d_t(c_b^*)$ matches the sign of $h_t(c_a^*)-h_t(c_b^*)$. This uses the full outer-holdout matrix rather than only its winner.

Unlike a terminal score, the outer-holdout matrix preserves selection regret and the ordering of every alternative.

We define \emph{endpoint breadth} as the number of endpoints for which one frozen intervention has positive outer-holdout effect, $h_t(c)>0$. Property-family breadth counts a family when at least one of its endpoints has positive effect. Compatibility is a separate post-search analysis. After the single-axis outer matrix was complete, we formed one deterministic Feature and Model assembly from the already frozen files without additional search or tuning. This assembly is excluded from the primary 9-of-10 selection result.

\section{Experimental Design}

\paragraph{Composition tasks.}
We use the four Matbench endpoints with composition input and experimental targets \citep{dunn2020matbench}. Experimental band gap and steel yield strength use MAE; metallicity and glass-forming ability use ROC-AUC. The baseline combines 132 Magpie descriptors \citep{ward2016magpie,ward2018matminer} with CatBoost \citep{prokhorenkova2018catboost}. Four campaigns search Feature, Model, Representation, and Data. Agent-curated candidate rows are filtered by reduced formula and source before sampling.

\paragraph{Structure tasks.}
Six MAE endpoints cover phonons (vibrational), bulk and shear moduli (elastic), refractive index (optical), and two-dimensional exfoliation plus perovskite formation energy (thermodynamic), spanning four property families. These are the remaining structure-input Matbench tasks with fewer than 20,000 examples. We omit the three Materials Project tasks with more than 100,000 examples so that every candidate can receive the same CPU-only five-fold evaluation budget. The campaign baseline adds an eight-descriptor density-and-symmetry block to Magpie, giving 140 inputs. Three campaigns search richer structure features, model and calibration changes, and composition embeddings. We evaluate the Data axis on composition tasks, where reduced formula supports exact overlap checks for external rows. Structure tasks instead hold the dataset fixed and compare Feature, Model, and Representation under matched crystal inputs and target definitions. All seven campaigns use the same 100-attempt stopping rule. One campaign recorded a 101st evaluation that completed at the stopping boundary, yielding 701 attempts without changing the frozen intervention. Two Structure-Model attempts terminated during execution before returning a score, leaving 699 scored attempts.

\paragraph{Outer-holdout protocol.}
We use the official Matbench fold-0 test as the outer holdout and build inner five-fold feedback only from its training partition. Outer-holdout labels are not loaded into an agent process. Once selection is complete, each frozen intervention is retrained on the complete outer training partition and scored on the outer holdout. We also replay frozen configurations over the five official Matbench folds for comparison with published references. Because four of those folds reuse examples from the search pool, this official-fold replay is post-selection benchmark context rather than a second independent test.

\paragraph{Agent and compute.}
DeepSeek-V4-Pro \citep{deepseekai2026v4} generated hypotheses and code changes under identical search contracts, budgets, and evaluator access for all axes. Experiments ran with Python 3.12 on a 32-vCPU AWS c7a.8xlarge instance; structure featurization and all model fitting were CPU-only. The analysis uses the complete set of successful and failed attempts, while headline comparisons use only the pre-defined baseline and frozen interventions.

\section{Results}

\begin{figure*}[t]
\centering
\includegraphics[width=\textwidth]{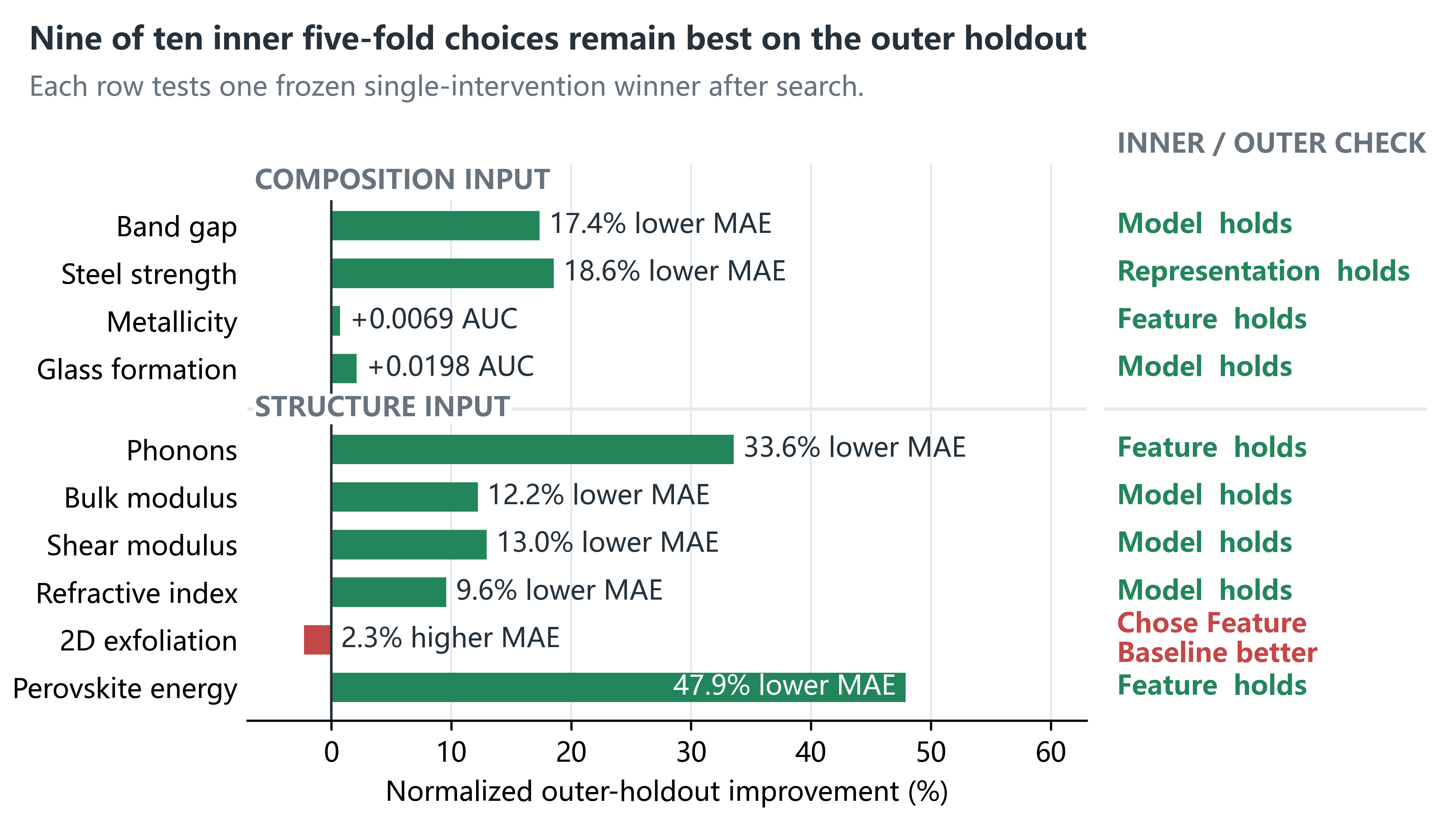}
\caption{Intervention selection on ten outer holdouts. Each row reports the effect of the frozen intervention chosen by inner five-fold feedback. The right column checks that choice against the best tested single intervention on the outer holdout; they agree on nine endpoints. Bars show $h_t$, the direction-corrected normalized outer-holdout improvement. Annotations show relative MAE reductions and absolute ROC-AUC changes.}
\label{fig:selection}
\end{figure*}

Table~\ref{tab:discoveries} lists the intervention returned by each campaign. Each entry describes frozen executable code rather than a label assigned after evaluation.

\begin{table}[t]
\centering
\small
\setlength{\tabcolsep}{1.8pt}
\renewcommand{\arraystretch}{1.08}
\begin{tabular}{@{}ll>{\raggedright\arraybackslash}p{0.34\columnwidth}rr@{}}
\toprule
\textbf{Input} & \textbf{Axis} & \textbf{Frozen output} & \shortstack{\textbf{Mean}\\\textbf{gain}} & \textbf{Breadth} \\
\midrule
\textbf{Composition} & Feature & Physics descriptors & $+6.6\%$ & $4/4$ \\
 & Model & Tree ensembles & $+6.0\%$ & $4/4$ \\
 & Repr. & Property-matched element pools & $+4.8\%$ & $3/4$ \\
 & Data & Screened candidate rows & $+0.2\%$ & $3/4$ \\
\addlinespace[2pt]
\textbf{Structure} & Feature & Coordination, geometry, strain, and bond chemistry & $+14.6\%$ & $5/6$ \\
 & Model & CatBoost--LightGBM ensemble & $+7.1\%$ & $5/6$ \\
 & Repr. & Additional composition embeddings & $-3.0\%$ & $3/6$ \\
\bottomrule
\end{tabular}
\caption{Frozen technical outputs, mean normalized outer-holdout improvement, and endpoint breadth. Breadth counts endpoints with positive outer-holdout effect. MAE uses relative error reduction and ROC-AUC uses relative gain against the corresponding baseline. Repr. denotes Representation.}
\label{tab:discoveries}
\end{table}

\subsection{Outer Evidence Confirms Nine, Rejects One}

Figure~\ref{fig:selection} gives the primary selection result. Inner five-fold feedback chooses four different intervention types across the ten endpoints. On composition tasks, it chooses a Model change for band gap and glass formation, a Representation change for steel strength, and a Feature change for metallicity. All four remain the best tested single intervention on their outer holdouts. Their effects include a 17.4\% MAE reduction for band gap, an 18.6\% reduction for steel strength, and absolute ROC-AUC gains of 0.0069 and 0.0198 for metallicity and glass formation.

On structure tasks, inner five-fold feedback selects the Feature intervention for phonons, two-dimensional exfoliation, and perovskite formation energy. It selects the Model intervention for both elastic moduli and refractive index. Five of these six choices remain best on the outer holdout. The exception is two-dimensional exfoliation. Its Feature intervention improves inner five-fold MAE from 42.85 to 38.79, but the 128-case outer holdout favors the baseline at 25.53 over 26.11. This endpoint has the smallest evaluation sample in the structure suite, with 128 outer cases and 101--102 cases per inner fold, and is the only task on which the two evidence sources disagree.

Selection regret gives a more graded view than the winner count. It is zero on nine endpoints and 2.279 percentage points on two-dimensional exfoliation, for a ten-task mean of 0.228 points. The full intervention ordering is also largely retained. The composition matrix contains 40 pairwise comparisons among the baseline and four axes, and the structure matrix contains 36 among the baseline and three axes. Inner five-fold and outer-holdout order agree on 67 pairs, reverse on eight, and tie on one, giving 89.3\% agreement among non-ties. Two-dimensional exfoliation is the only negative-correlation task and accounts for four of the eight reversals. The official-fold replay preserves all ten choices; for metallicity, Feature is the best single axis on every replay fold.

Because the alternatives emerge through repeated reuse of inner feedback rather than from a fixed shortlist, we compare the observed winner count with an exact blocked null. It applies one shared relabeling of inner-feedback choices to outer-holdout interventions within each input regime, preserving each frozen intervention's identity across tasks. The baseline plus four composition axes admit $5!$ relabelings, and the baseline plus three structure axes admit $4!$, giving $5!\times4!=2{,}880$ mappings. Chance therefore matches 2.3 outer winners in expectation, and only 4 mappings match at least nine ($P=0.0014$). Together, winner count, regret, and ordering measure decision reliability. Nine of ten selections and 89.3\% of non-tied orderings survive outside the loop; median task-wise Spearman correlation is 0.95.

\paragraph{Search dynamics.}
Of 699 scored attempts, 131 improve on their parent. Composition yields 33 Model, 21 Feature, 13 Representation, and 6 Data improvements; structure yields 31 Feature, 17 Model, and 10 Representation improvements. Choices stabilize by attempt 50, while final code appears at attempts 47--100 (median 96).

\subsection{Intervention Hierarchies Depend on Input}

\begin{figure*}[t]
\centering
\includegraphics[width=\textwidth]{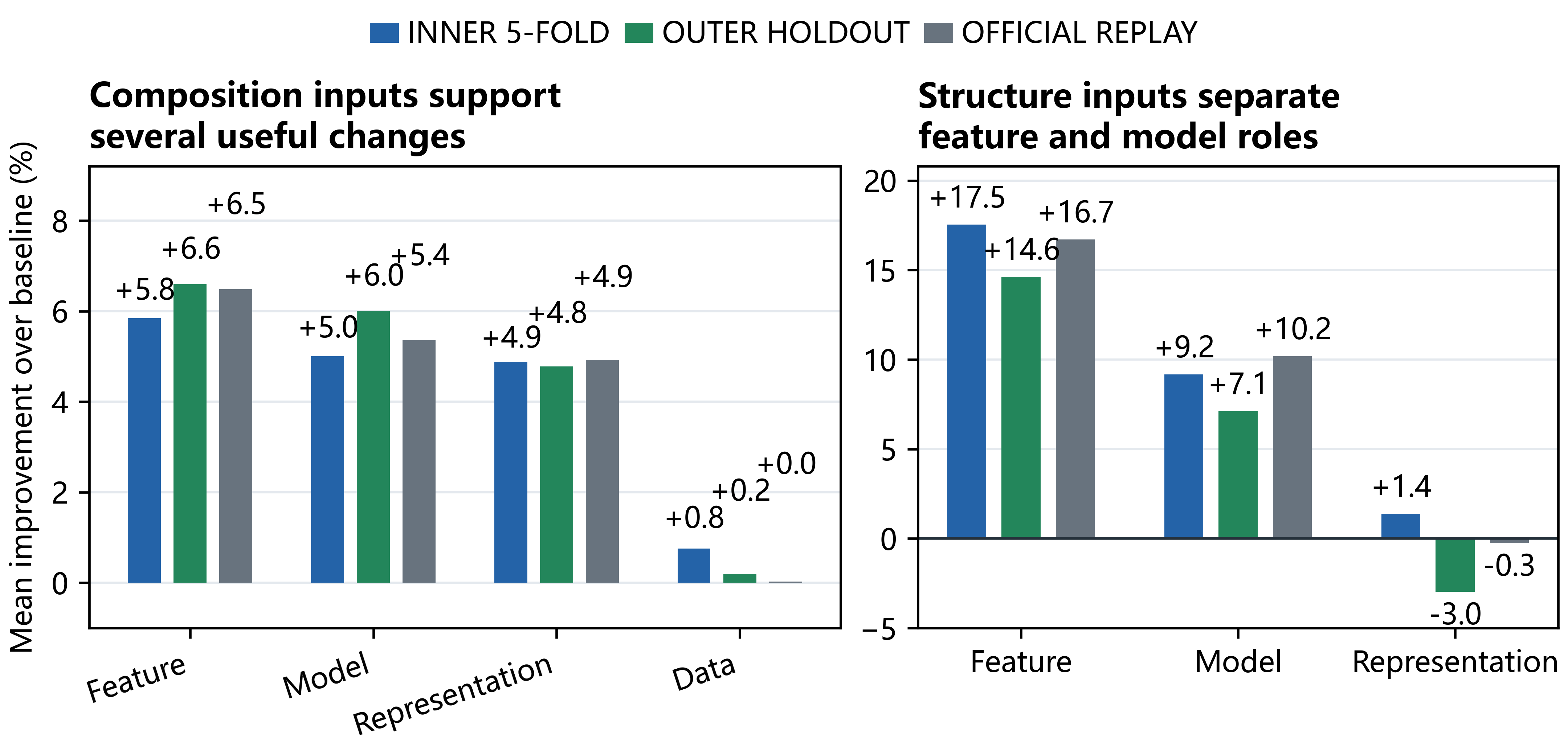}
\caption{Intervention hierarchies depend on input. Composition supports several useful changes, while structure separates geometry features and models. Data is tested only for composition, which has screened external rows. Structure Representation gains on inner feedback but loses on the outer holdout. Official replay gives post-selection context.}
\label{fig:regimes}
\end{figure*}

Figure~\ref{fig:regimes} aggregates the frozen intervention from each axis relative to its baseline. On the four composition tasks, Feature, Model, and Representation interventions improve the outer-holdout mean by 6.6\%, 6.0\%, and 4.8\%, respectively. The Data intervention adds only 0.2\%. No single axis dominates composition prediction.

A different hierarchy appears on the structure-informed tasks. Across the six structure tasks, richer Feature code lowers mean outer-holdout MAE by 14.6\%, and Model code lowers it by 7.1\%. Added composition embeddings increase error by 3.0\%. These are the two strongest tested axes in this setting. Geometry features reduce phonon MAE by 33.6\% and perovskite formation-energy MAE by 47.9\%. Model and calibration changes reduce bulk-modulus, shear-modulus, and refractive-index errors by 12.2\%, 13.0\%, and 9.6\%.

\paragraph{Outer evidence rejects the selected Representation.}
The structure Representation intervention gains 1.4\% under inner five-fold feedback but raises outer-holdout error by 3.0\%. Losses concentrate on phonons and two-dimensional exfoliation, the two smallest structure datasets, where it adds formula encodings but no local geometry. The controlled comparison below holds the task and estimator fixed.

\paragraph{Input type, not input dimension.}
Holding the phonon and bulk-modulus endpoints, CatBoost estimator, and official five-fold protocol fixed, we compare three input blocks. The 132-dimensional Magpie baseline has errors of 66.05 cm$^{-1}$ and 0.0837 in $\log_{10}(\mathrm{GPa})$. A 348-dimensional composition encoding formed by adding mat2vec and MEGNet element embeddings changes them to 79.26 cm$^{-1}$ and 0.0847. Magpie with nine light structure descriptors, including packing efficiency, instead has 141 inputs and errors of 54.98 cm$^{-1}$ and 0.0655, reductions of 17\% and 22\%. More formula-derived capacity does not replace direct structural information.

The 100-attempt Data campaign curates, filters, and recombines candidate rows under strict formula and source checks. Its best intervention improves mean outer-holdout performance by 0.2\%, compared with 4.8--6.6\% for the other composition axes. Data curation is therefore the least productive tested action in this regime. Axis isolation places this focus of data-centric AI \citep{zha2025datacentric} on the same quantitative footing as model and representation changes.

\paragraph{Reusable interventions.}
The frozen interventions are inspectable code. Composition interventions route tasks among element identity, alloy descriptors, compact Magpie statistics, tuned estimators, and tree ensembles. Structure Feature combines coordination, bond geometry, strain, oxidation state, electronegativity, and $d$-electron statistics; Model combines CatBoost and LightGBM with fitted MAE weights \citep{prokhorenkova2018catboost,ke2017lightgbm}. Each file-level change can be replayed, ablated, and combined.

The results suggest an allocation principle based on information rather than model brand. Composition descriptors, element representations, and tree models reorganize the same formula evidence, so several routes remain productive. Once coordinates are present, coordination, strain, and bond chemistry supply variables that the tested larger formula encoding cannot reconstruct. The 216 added composition dimensions and the nine geometry descriptors quantify this distinction. The useful next action depends on whether the task lacks information or lacks a suitable mapping of information already available.

\subsection{Feature and Model Changes Remain Compatible}

The frozen structure Feature and Model interventions each lower error on five of six outer holdouts and reach all four property families, but they specialize. Feature leads on phonons and perovskite formation energy, while Model leads on elastic and refractive properties. The Model intervention was selected for the 140-input baseline, whereas Feature changes the input distribution and the variables available to every tree. The two changes could therefore be redundant or interfere even when each helps alone.

\begin{table}[t]
\centering
\small
\setlength{\tabcolsep}{2.5pt}
\begin{tabular}{lcrrr}
\toprule
\textbf{Task} & \textbf{Best axis} & \textbf{Single (\%)} & \textbf{F+M (\%)} & \textbf{Gain (pp)} \\
\midrule
Phonons & F & 33.57 & 48.38 & $+14.81$ \\
Bulk modulus & M & 12.21 & 15.80 & $+3.59$ \\
Shear modulus & M & 12.96 & 16.96 & $+4.00$ \\
Refractive index & M & 9.58 & 12.48 & $+2.90$ \\
2D exfoliation & F & $-2.28$ & 0.94 & $+3.22$ \\
Perovskite energy & F & 47.90 & 62.95 & $+15.05$ \\
\midrule
Mean & & 19.0 & \textbf{26.3} & $\mathbf{+7.3}$ \\
\bottomrule
\end{tabular}
\caption{Outer-holdout MAE reduction in the deterministic F+M compatibility test. ``Single'' is the best single axis without the baseline; Gain is the improvement over that component. Official five-fold gap closure is 56.6--88.2\% (mean 71.5\%) \citep{chen2019megnet,congn,debreuck2021modnet,matbench_archive}.}
\label{tab:compose}
\end{table}

The complementary profiles in the completed single-axis matrix motivate one deterministic compatibility test without further search or tuning. Their union improves all six structure tasks and reduces mean outer-holdout MAE by 26.3\%, compared with 19.0\% for the best single intervention and 14.6\% for Feature alone. Official-fold replay preserves this ordering at 27.0\% versus 20.3\%, closes 71.5\% of the gap from the CPU tabular baseline to the best archived entry, and places refractive-index MAE within 2.2\% of MODNet, without graph-neural-network training.

\FloatBarrier
\section{Conclusion}

We introduced intervention-centered Auto Research, which independently searches declared axes, freezes their selected code, and evaluates every intervention in one outer-holdout matrix. The method makes the research decision, rather than an entangled terminal pipeline, the unit of evidence and measures selection, ordering, breadth, and compatibility.

Across 701 attempts and ten Matbench endpoints, inner feedback selected the best tested intervention on nine outer holdouts, with 0.228 percentage-point mean regret and 89.3\% ordering agreement. The matrix also rejected a structure Representation intervention that gained 1.4\% inside the loop but lost 3.0\% outside it.

The materials evidence identifies an information-dependent hierarchy. Composition-only tasks support several axes, while structure tasks favor geometry Feature and complementary Model changes. Holding endpoints, CatBoost, and official folds fixed, nine structure descriptors reduce errors by 17\% and 22\%, whereas 216 composition dimensions do not help. A deterministic Feature and Model union improves all six outer holdouts, raising the mean from 19.0\% to 26.3\%; on official replay it closes 71.5\% of the archive gap.

Decision-level validation makes adaptive agent search reusable wherever alternatives meet fixed evaluators.

\paragraph{Generative AI use.}
Generative AI assisted with language and figure editing; the authors remain responsible for all content.

\bibliography{references}

\end{document}